\documentclass[useAMS,usenatbib]{mn2e}

\usepackage{color}
\usepackage{units}
\usepackage{amsmath}
\usepackage{graphicx}

\newcommand{\Ms}{M$_{\odot}$}
\newcommand{\Ls}{L$_{\odot}$}
\newcommand{\Rvir}{{\ensuremath{R_{200}}}}

\def\aaps{A\&AS}
\def\aap{A\&A}
\def\apj{ApJ}
\def\apjl{ApJL}
\def\apjs{ApJS}
\def\araa{ARA\&A}
\def\jqsrt{JQSRT}
\def\jpcrd{JPCRD}
\def\mnras{MNRAS}
\def\nat{Nature}
\def\optcomm{Opt.~Comm.}
\def\ssr{Space~Sci.~Rev.}

\title[Lyman-Werner UV Escape Fractions]{Lyman-Werner UV Escape Fractions from
Primordial Halos}

\author[Schauer et al.]{Anna T. P. Schauer$^{1}
$\thanks{E-mail:schauer@uni-heidelberg.de},
Daniel J. Whalen$^{1,2}$, Simon C. O. Glover$^{1}$, Ralf S. Klessen$^{1}$\\$^{1}$
Universit\"at Heidelberg, Zentrum f\"ur Astronomie, Institut f\"ur Theoretische
Astrophysik, Albert-Ueberle-Str. 2, 69120 Heidelberg, Germany\\
$^{2}$Institute for Cosmology and Gravitation, University of Portsmouth, Portsmouth, PO1 3FX, UK}

\begin{document}

\date{Accepted 2015 September 10.  Received 2015 August 19; in original form 2015 May 21.}

\pagerange{\pageref{firstpage}--\pageref{lastpage}} \pubyear{2002}

\maketitle

\label{firstpage}

\begin{abstract}

Population III stars can regulate star formation in the primordial Universe in several ways.
They can ionize nearby halos, and even if their ionizing photons are trapped by their own
halos, their Lyman-Werner (LW) photons can still escape and destroy H$_2$ in other halos,
preventing them from cooling and forming stars.  LW escape fractions are thus a key
parameter in cosmological simulations of early reionization and star formation but have not
yet been parametrized for realistic halos by halo or stellar mass.  To do so, we perform
radiation hydrodynamical simulations of LW UV escape from 9--120 \Ms\ Pop III stars in
$10^5$ to $10^7$ \Ms\ halos with ZEUS-MP.  We find that photons in the LW lines
(i.e.\ those responsible for destroying H$_{2}$ in nearby systems) 
have escape fractions ranging from 0\% to 85\%.  No LW photons escape
the most massive halo in our sample, even from the most massive star.  Escape fractions
for photons elsewhere in the 11.18--13.6~eV energy range, which can be redshifted
into the LW lines at cosmological distances,
are generally much higher, being above 60\% for all but the least massive stars in the most
massive halos.  We find that shielding of H$_2$ by neutral hydrogen, which has been
neglected in most studies to date, produces escape fractions that are up to a factor of
three smaller than those predicted by H$_2$ self-shielding alone.

\end{abstract}


\section{Introduction}

The first stars form at $z \sim 20$-30, or about 200 Myr after the big bang.  Early numerical
simulations suggested that Pop III stars are very massive, 100 - 500 \Ms, and form in
isolation, one per halo \citep{bcl99,bcl02,abn00,abn02,nu01,glover05}. More recent models
indicate that most Pop III stars form in binaries \citep{turk09} or in small clusters 
(\citealt{stacy10}; \citeauthor{clark11}~2011a,b; \citealt{get11,get12,sm11,
dgck13,glov13,sb14}) with a wide
range of masses, all the way from the sub-solar regime to $\sim$$10^3\,$M$_\odot$.
These calculations suggest that the final mass of a star is either limited by
dynamical processes (\citeauthor{cgkb11}~2011a; \citealt{get12,shogk12})
or by radiative feedback \citep{tm08,hos11,hos12,stacy12,hir14}. 

If any Pop III stars formed with masses below $\sim$$0.8\,$M$_\odot$, they will 
have survived until the present day,
and they may therefore be detectable in surveys targeting extremely metal-poor stars in
the Galactic bulge and halo as well as in nearby satellite galaxies 
(see the reviews by \citealt{bc05}, or \citealt{frebel10}). This could potentially
allow us to set limits on the low mass end of the primordial stellar initial mass function
\citep[IMF, see e.g.][]{tum06,ssa07,hb15}.
Unfortunately, the direct detection of Pop III stars in the high redshift
Universe is unlikely even with future 30 meter telescopes \citep[although see][]{rz10}.
As massive stars are very short lived, the upper end of the Pop III IMF is thus more
difficult to constrain. This can only be done indirectly, for example by observing and
analyzing the supernovae (SNe) that mark the end of the lives of very massive stars. These may indeed
be visible to the {\it James Webb Space Telescope} ({\it
JWST}), the Thirty-Meter Telescope (TMT), the Extremely Large Telescope (ELT) and the
{\it Giant Magellan Telescope} ({\it GMT}) \citep[e.g.,][]{wet12b,wet12c,wet13d,smidt14a}.
Another promising possibility is again Galactic archeology. The comparison of
the nucleosynthetic yields of Pop III SNe to the chemical abundances of extremely
metal-poor stars observed in our Milky Way suggests that many Pop III stars
may had masses in the range $15 - 40\,$\Ms\ \citep[e.g.,][]{bc05,fet05,fjb08,jet10}.

In this paper, we focus on the high-mass end of the Pop III IMF. Massive Pop III stars had the
capability to profoundly transform their environment due to the strong
radiative, chemical, and mechanical feedback they provide. They can evaporate the gas from
their dark matter (DM) halos, engulf nearby halos with both
ionizing and LW band UV radiation \citep{wan04,abs06,awb07,wet08b}, and  alter their
composition with the metal they produced and expelled. LW photons lie in
a series of discrete lines located in the energy range 11.18 -- 13.6 eV
(corresponding to a wavelength range of 1110$\,$\AA\ -- 912$\,$\AA\ )
and can travel great distances in the primeval
universe because they lie below the ionization threshold of atomic hydrogen. They can delay or suppress
star formation in nearby halos because they photodissociate H$_2$ and prevent the gas
from cooling \citep[see e.g.][]{hrl97,har00,gb01,gb03,met01,su06,su07,on07,wa08b}.  How Pop III stars
regulate subsequent star formation is central to the rise of stellar populations in the first
galaxies \citep{cf05,jlj09,get10,jeon11,pmb12,wise12}.  This begins with understanding how
the strength of the LW background changes over time, which depends on the masses of
Pop III stars and the halos in which they reside.

Over the past two decades, many studies have examined the influence of LW and ionizing
radiation on the formation of stars at high redshift. They found  that a
photodissociating background cannot halt Pop III star formation but can delay 
it \cite[e.g.,][]{met01,wa07, on07,saf12}.
An early X-ray background created by accretion onto stellar remnants and by emission
from supernovae can catalyze
additional H$_2$ formation that offsets the effects of LW radiation to some extent, but
radiative feedback overall is dominated by the LW background \citep{gb03,mba03}. The evolution
of this background over time in general seems to play only a minor role if it changes
less rapidly with redshift $z$ than $10^{-z/5}$ \citep{vis14}. 
Simulations suggest that Pop III stars forming in halos illuminated by significant
LW backgrounds will be more massive because their host halos must grow to larger masses
before beginning to cool \citep{on07,hir15,latif14}.  When they do begin to cool, collapse rates at
their centers are higher and lead to more massive stars.

The effect of a single Pop III star on a nearby halo has been examined in much greater
detail than in large cosmological boxes.  Radiation hydrodynamical simulations by
\citet{su06} find that star formation in the vicinity of a source star is possible if the star
forming region exceeds a density threshold \citep[see also][]{gb01}.
Other studies in this vein show that star
formation can be either promoted or suppressed depending on the mass of the
halo, the mass of the star, and
the proximity of the halo to the star (\citealp{su07,wet08b,wet10}; \citealp[see also][]{il04,il05,
hus09,suh09}).

In some cases the H$\,${\sc{ii}} region of a Pop III star may fail to break out from a halo because of
large central gas
densities.  In such cases, LW photons might still exit the halo because their energies lie
below the ionization limit of hydrogen, even if H$_2$ self-shields against this flux to some
degree.  Most simulations of radiative feedback with Pop III stars in cosmological boxes
simply assume uniform LW backgrounds.  A parametrization of LW escape fractions from
primordial halos as a function of halo and stellar mass could provide the strength of this
background from first principles in future simulations. However, almost no work has
been done to produce such a parametrization. In the only previous study on this topic of
which we are aware,  \citet{ket04} examined LW escape
fractions from Pop III star-forming halos with a \cite{nfw97} 
radial density profile, varying the mass of the
halo and central star.

In an effort to improve on this previous study, we perform radiation hydrodynamical calculations of LW escape from Pop
III minihalos with the ZEUS-MP code.  We then post-process the simulations with semi-analytical
methods and calculate escape fractions in
two limits, the near-field and the far-field. In the near-field case -- the only
scenario considered by \citet{ket04} -- we compute the escape fraction of photons
in the LW lines themselves. This is the value that is relevant if we are interested in the
effect of radiation from the halo on H$_{2}$ in its immediate vicinity.
In the far-field case, on the other hand, we are interested in the total fraction of the
photons lying between 13.6 eV and 11.2 eV that can escape 
from the halo. This
is the important quantity if we are interested in the effect of radiation from the halo
on H$_{2}$ located at cosmological distances.

In both cases, we account not only for H$_{2}$ self-shielding but also for shielding
from the Lyman series lines of atomic hydrogen, which were not considered by
\citet{ket04}. In some circumstances, this can significantly affect the LW escape
fraction.

Our paper is structured as follows. In Section 2 we describe our
numerical models.  In Section 3 we tabulate LW escape fractions by halo and stellar
masses and we conclude in Section 4.

\section{Method}

To calculate LW escape fractions for a given star and halo we first evolve the H$\,${\sc{ii}} region
of the star with the radiation hydrodynamics simulation code ZEUS-MP.
We then post-process the profiles for the ionization front (I-front) and the
surrounding halo with semi-analytic calculations to determine how many LW photons exit
the halo.  LW escape fractions in both the near-field and far-field approximations are
considered, as described below.

\subsection{ZEUS-MP}

ZEUS-MP is an astrophysical radiation-hydrodynamics code that self-consistently couples
photon-conserving raytracing UV transport and nonequilibrium primordial gas chemistry to
gas dynamics to evolve cosmological I-fronts \citep{wn06,wn08b,wn08a}. We evolve mass
fractions for H, H$^{+}$, He, He$^{+}$, He$^{2+}$, H$^{-}$, H$^{+}_{2}$, H$_{2}$, and e$
^{-}$ with nine additional continuity equations and the nonequilibrium rate equations of
\citet{anet97} in which the species are assumed to share a common velocity distribution.
Mass and charge conservation, which are not formally guaranteed by either the network or
advection steps, are enforced at every update of the reaction network.  Heating and
cooling due to photoionization and chemistry are coupled to the gas energy density with an
isochoric update that is operator-split from
the fluid equations.  Cooling due to
collisional ionization and excitation of H and He, recombinations of H and He, inverse
Compton scattering (IC) from the CMB, bremsstrahlung emission, and H$_2$ cooling are
all included in our models. The chemical and cooling rate coefficients are the same
as those used in \citet{anet97}, with one important exception: we use case B rates to
describe the recombination of hydrogen and helium, rather than case A as used in
\citet{anet97}.

We use 120 energy bins in our photon-conserving UV transport scheme, 40 bins that are
uniform in energy from 0.755 to 13.6 eV and 80 bins that are logarithmically spaced from
13.6 eV to 90 eV.  We normalize photon rates in each bin by the time-averaged ionizing
photon rates and surface temperatures for Pop III stars from Tables 3 and 4 of \citet{
schae02}.  The radiative reactions in our simulations are listed in Table~1 of \citet{wn08b},
and the momentum imparted to the gas by ionizations of H and He is included in the photon
transport.  Photon conservation is not used to calculate H$_2$ photodissociation rates.
They are derived along radial rays from the star with the self-shielding functions of 
\citet[][hereafter DB96]{db96} modified for thermal broadening as a proxy for the effects of gas motion
\citep[equations 9 and 10 in][]{wn08b}, with $r^{-2}$ attenuation taken into account.

Two-photon emission from recombining He$^{+}$ and He$^{2+}$ can produce
photons capable of photo-dissociating H$_2$. However, we do not account for the effects of
this nebular emission, as it is unimportant in comparison to the effects of the direct
stellar emission.

\subsection{Halo Models}

\begin{figure*}
\includegraphics{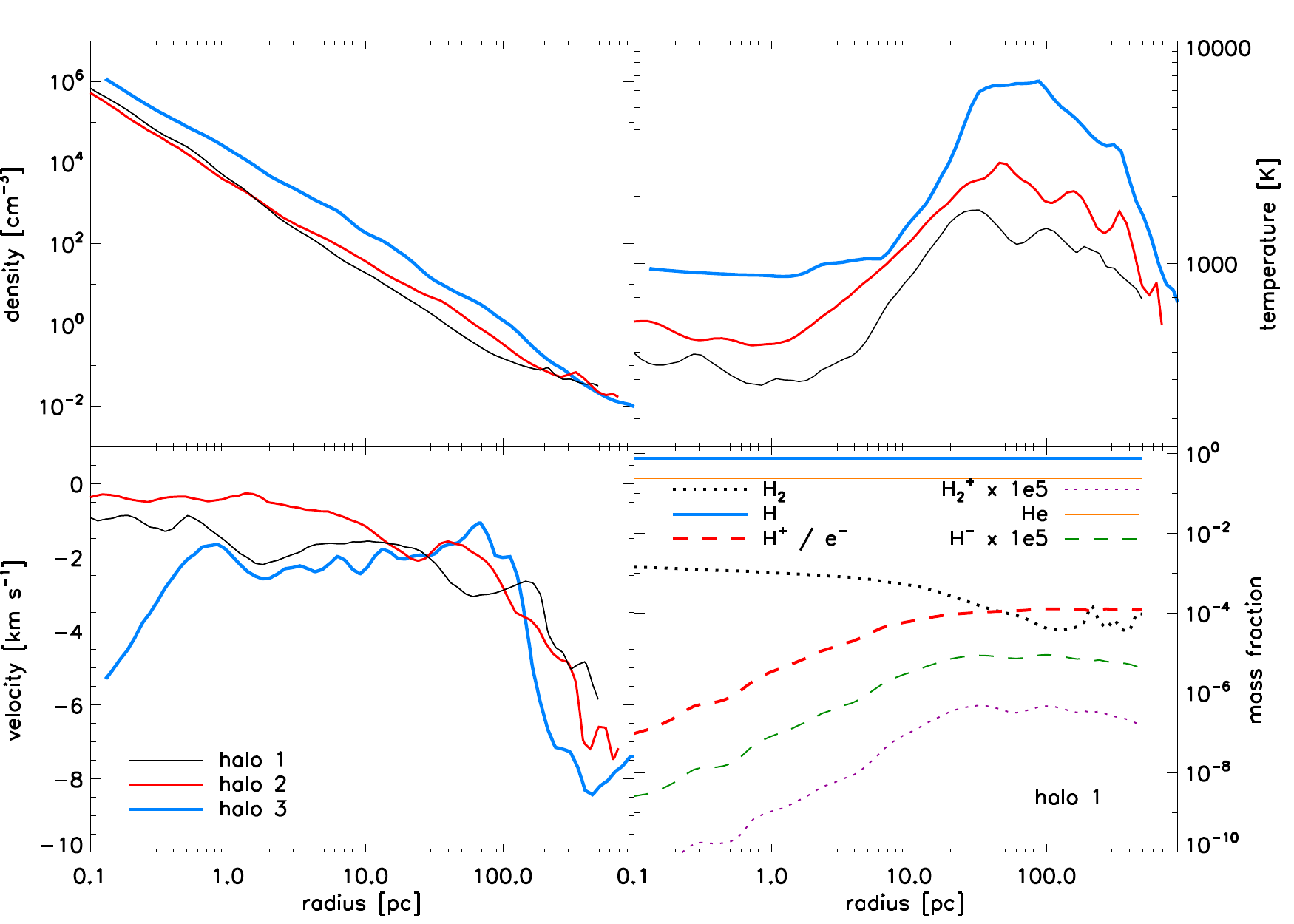}
\caption{Initial densities (upper left), temperatures (upper right), radial velocities (lower left) and 
species mass fractions (lower right) for the three halos.}
\label{fig:haloics}
\end{figure*}

We adopt the one-dimensional (1D) spherically-averaged halo profiles used in \citet{
wet10}, which were based on the results of cosmological simulations carried
out with the Enzo adaptive 
mesh refinement (AMR) code \citep{bryan14}.  
Their masses are 6.9 $\times$ 10$^5$ \Ms, 2.1 $\times$ 10$^6$ \Ms and 1.2 $\times$ 10$^7$ \Ms.
This corresponds to the range of halo masses in which Pop III stars are expected to form via 
H$_2$ cooling, and all three halos form at z $\sim$ 20. Densities, velocities, temperatures and mass 
fractions for all nine primordial species are mapped from these profiles onto a 1D 
spherical grid in ZEUS-MP.  We summarize the properties of these halos in 
Table~\ref{tbl:r200} and plot their initial densities, velocities, temperatures, and species mass 
fractions in Figure~\ref{fig:haloics}.  

The gas densities have nearly a power-law profile with slopes around -2.1, with the most massive 
halo having the highest densities. Temperatures vary from a few hundred K to several thousand K 
and generally increase with halo mass. This is to be expected since the virial temperature of a halo
scales with the halo mass as $T_{\rm vir} \propto M^{2/3}$.
Virial shocks are visible in all three profiles at $\sim$ 100 pc, where infall velocities 
abruptly decrease as accretion flows crash into increasingly dense regions of the halo.  
Shock heating at these radii is also evident in the temperature profiles, and is almost
strong enough to collisionally ionize atomic hydrogen in halo 3.

ZEUS-MP does not evolve DM particles as in cosmological codes such as Enzo or GADGET; instead, 
an additional gravitational potential is implemented as a proxy for the DM potential of the halo.
We take its potential to be that required 
to keep the baryons in hydrostatic equilibrium on the grid.  
The mass associated with this 
potential is nearly the same as that of the halo. We interpolate this precomputed potential 
onto the grid at the beginning of the run but it does not evolve thereafter.  Taking the 
potential to be static is a reasonable approximation because merger and accretion 
timescales at $z \sim$ 20 are on the order of 20 Myr, significantly longer than the lifetimes
of most of the stars that we consider in this study (see Table~\ref{tbl:props}). We therefore do not
expect the dark matter distribution to evolve much over the lifetime of the star.  

\subsection{H{\sc{ii}} Region Simulation Setup}

We center the Pop III star at the origin of a 1D spherically-symmetric grid that
has 500 ratioed zones in radius. The 
inner boundaries are at 2.0 $\times$ 10$^{17}$ cm (halos 1 and 2) or 4.0 $\times$ 10$^{
17}$ cm (halo 3) and the outer boundaries are at slightly more than twice the virial radius 
of the halo.  Reflecting and outflow conditions are imposed on the inner and outer 
boundaries, respectively. Densities, energies, velocities and species mass fractions from 
our Enzo profiles are mapped onto the ZEUS-MP grid with a simple linear interpolation of 
the logarithm of the given variable. We consider Pop III stars with masses 9, 15, 25, 40, 60, 
80 and 120 \Ms\ in all three halos, leading to a total of 21 models. 
The properties of these stars are listed in 
Table 2 \citep{schae02}.  Each simulation is run out to the end of the life of the star.

In two of the 21 simulations (the 15 \Ms\ star in halos 1 and 2) the I-front is confined to 
small radii that require a finer grid in order to be resolved.  We therefore use a grid of  
1000 zones with two contiguous blocks: a very finely spaced uniform grid with 600 cells 
to resolve the I-front followed by a ratioed grid with 400 cells that again extends to 
twice the virial radius of the halo.  We ensure that the length of the innermost zone of 
the outer block is within 20\% of that of the outermost zone of the inner block to avoid 
spurious reflections of shocks at the interface of the two blocks. 

\subsection{LW Escape Fractions}
Molecular hydrogen has no dipole moment and is therefore destroyed in a two-step 
photodissociation process, the Solomon-process \citep{sw67}: 
$\mathrm{H}_2 + \gamma \rightarrow \mathrm{H}_2^\star \rightarrow 2 \mathrm{H}$. 
Incident radiation can excite the molecule from the ground state to an electronically
excited state. Two of these excited states, $\mathrm{B}^1\Sigma^+_\mathrm{u}$ (known
as the Lyman state) and $\mathrm{C}^1\Pi_\mathrm{u}$ (the Werner state), are separated
from the electronic ground state by less than 13.6~eV, and transitions to these states can
therefore be brought about by photons with energies below the Lyman limit of atomic
hydrogen. Within the Lyman and Werner states, a variety of bound rotational or vibrational
levels are accessible, and so transitions from the ground state to the Lyman or Werner states
occur through a series of discrete lines, known as the Lyman-Werner band system, or
simply as the Lyman-Werner lines. Once excited by a photon in one of these lines (a LW
photon), the H$_{2}$ molecule remains in the Lyman or Werner state for only a very short
time $\Delta t \sim 10^{-8} \: {\rm s}$, before decaying back to the electronic ground state.
Most of the time, the H$_2$ molecule decays to a bound ro-vibrational level in the ground
state. However, a small fraction of the time, the decay occurs to the vibrational continuum,
resulting in the dissociation of the H$_2$ molecule. The dissociation probability depends
on the details of the incident spectrum and the density and temperature of the gas (which
fix the initial rotational and vibrational level populations), but is typically around 15\% (DB96).
The remaining 85\% of LW photon absorptions result in decay back to a bound state.
In a small fraction of cases, this results in a photon with the same energy as the original
LW photon (albeit with a random direction). However, the majority of the time, the 
re-emitted photon has too small an energy to bring about photodissociation \citep[for more
details, see the discussion in Section 3.4 of]{gb01}.
  
When the H$_{2}$ column density is large, the gas becomes optically 
thick in the LW lines. This can prevent radiation from an external source from reaching 
the center of a halo or LW flux from the center of a halo from escaping it. The gas 
therefore self-shields against LW radiation. In addition, LW photons can also be
absorbed by most of the Lyman series lines of atomic hydrogen.\footnote{The exception
is Lyman-$\alpha$, which is located at too low an energy.} However, because the frequencies
of the LW lines of H$_{2}$ do not coincide particularly closely with those of the Lyman series
lines of H, this effect only becomes important when the atomic hydrogen column density
is very high, $N_{\rm H} \simeq 10^{23} \: {\rm cm^{-2}}$ or above, so that the Lyman series
lines are strongly Lorentz-broadened \citep{wh11}.

In the immediate vicinity of a halo, the photons responsible for photodissociating H$_{2}$
are those emitted in the LW lines in the rest-frame of the halo. The absorption of these photons
by H$_{2}$ and H within the halo can be conveniently parameterized by a simple self-shielding
function (see e.g.\ DB96; \citealt{wh11}, hereafter simply WH11). 
The escape fraction of photons in this limit, which we term
the near-field limit, is then simply given by the value of this self-shielding function evaluated at the
virial radius of the halo.
At larger, cosmological distances, the LW 
flux is redshifted by the expanding Universe or Doppler shifted due to relative velocities between 
halos. As a consequence, the LW lines in the source frame may not coincide with LW absorption 
lines in the local frame of a halo. A simple estimate of the escape fraction in this limit, which we
term the far-field limit, is given by the fraction of the LW 
range 11.2 -- 13.6 eV not lying inside the combined equivalent width of the absorption 
lines. In the sections below, we describe in more detail how we compute the 
LW escape fractions in both limits.  

\subsubsection{Near-Field Limit}

A LW photon escapes a halo if it does not photodissociate an H$_2$ molecule or is not 
absorbed and re-emitted as lower-energy photons. The LW escape fraction, $f_\mathrm{
esc}$, can therefore be equated to the fraction of H$_2$ molecules that are shielded from 
either process. The factors by which H$_2$ is shielded from LW photons by other H$_2$ 
molecules and by H atoms are $f_\mathrm{shield}^{\mathrm{H}_2}$ and $f_\mathrm{
shield}^\mathrm{H}$, respectively.  Both factors account for all processes by which LW 
photons are absorbed, not just those that result in a photodissociation.  The total factor 
by which H$_2$ is shielded can be taken to be the product of these two factors 
(see the commentary on equation 13 in WH11),
\begin{equation}
f_\mathrm{esc} = f_\mathrm{shield}^{\mathrm{H}_2} f_\mathrm{shield}^\mathrm{H}.
\end{equation}
To construct $f_\mathrm{shield}^{\mathrm{H}_2}$ we first determine the column density of H$_2$, 
\begin{equation}
N_{\mathrm{H}_2} = \sum^{r_i \le \Rvir}_{i=1} 
                            {n_i  \, \chi_{\mathrm{H}_2,i} \, (r_i -r_{i-1}) },
\end{equation}
where $n_i$ is the number density of all particles and $r_i$ is the outer radius of 
the $i^{th}$ cell, where $r_0$ is zero.
$\chi_{\mathrm{H}_2,i} = N_{\mathrm{H}_2,i} / N_{\mathrm{total},i}$ is the 
H$_2$ particle fraction with $N_{\mathrm{H}_2,i}$ the number of $\mathrm{H}_2$ molecules 
and $N_{\mathrm{total},i}$ the total number of particles in bin $i$. 
The sum extends out to the virial radius of the halo,
%
\begin{equation}
\Rvir = R(\rho = 200 \, \Omega_\mathrm{b,0} \, \rho_\mathrm{crit}), 
\end{equation}
where $\rho$ is the gas density and
$\rho_\mathrm{crit} = 3 H^2(z=20)/(8\pi G) \sim 2.349 \times 10^{-26} \mathrm{g}\ \mathrm{cm}^{-3}$ 
is the critical overdensity of the Universe at $z=20$. The virial radii are listed 
for all three halos in Table \ref{tbl:r200}.
 
\begin{table}
\centering
\begin{tabular}{c|cc}
halo & $\Rvir$ (pc) & mass (\Ms) \\
\hline
1 & 256.8 & $6.9 \times 10^5$ \\
2 & 339.6 & $2.1 \times 10^6$ \\
3 & 495.2 & $1.2 \times 10^7$ \\
\end{tabular}
\caption[r200]{Virial radii and total (dark matter $+$ baryon) masses of the three halos.}
\label{tbl:r200}
\end{table}

\begin{table}
\centering
\begin{tabular}{cccc}
$M$ (\Ms)  & lifetime (Myr) & log $L$ (\Ls) & log $T_\mathrm{eff}$ (K) \\
\hline
9    & 20.2  & 3.709 & 4.622  \\
15   & 10.4  & 4.324 & 4.759  \\
25   & 6.46  & 4.890 & 4.850  \\
40   & 3.86  & 5.420 & 4.900  \\
60   & 3.46  & 5.715 & 4.943  \\
80   & 3.01  & 5.947 & 4.970  \\
120  & 2.52  & 6.243 & 4.981  \\
\end{tabular}
\caption{Pop III stellar properties \citep{schae02}.}
\label{tbl:props}
\end{table}

The Doppler broadening of lines must be taken into account in $f_\mathrm{shield}^
{\mathrm{H}_2}$. The thermal component of the Doppler broadening parameter, $b_
\mathrm{D,T}$, is associated with the Maxwellian velocity distribution of the atoms 
and molecules,
\begin{equation}
P_v(v) \mathrm{d}v = \sqrt{\frac{m}{2 \pi k_B T}} \exp \left(-\frac{m v^2}{2 k_B T} \right).
\end{equation} 
From DB96, the Doppler broadening parameter is defined by $b = \mathrm{
FWHM}/(4 \ln{2})^{1/2}$, where FWHM is the full-width half maximum of $P_v(v)$ and 
is related to the standard deviation, $\sigma$, of $P_v(v)$ by $\mathrm{FWHM} = 2 
\sqrt{2 \ln{2}} \sigma$.  This yields
\begin{equation}
b_\mathrm{D,T} = \sqrt{ \frac{2 k_B T}{m_{\mathrm{H}_2}} },
\label{eq:bdoppler}
\end{equation}
where $k_B$ is the Boltzmann constant
and $m_{\mathrm{H}_2}$ is the mass of an H$_2$ molecule.  The temperature $T$ in equation 
(\ref{eq:bdoppler}) is the H$_2$-weighted mean temperature $T_\mathrm{eff}$ of the 
gas
\begin{align}
T_\mathrm{eff} &= \frac
{ \sum^{r_i \le \Rvir}_{i=1} {n_i \, \chi_{\mathrm{H}_2,i} \, T_i \, V_i} }
{ \sum^{r_i \le \Rvir}_{i=1} {n_i \, \chi_{\mathrm{H}_2,i} \, V_i} } \nonumber \\
               &= \frac
{ \sum^{r_i \le \Rvir}_{i=1} {n_i \, \chi_{\mathrm{H}_2,i} \, T_i 
                                  \, \frac{4\pi}{3}(r_i^3-r_{i-1}^3)} }
{ \sum^{r_i \le \Rvir}_{i=1} {n_i \, \chi_{\mathrm{H}_2,i} 
                                  \, \frac{4\pi}{3}(r_i^3-r_{i-1}^3)} },
\end{align}
where $T_i$ is the temperature and $V_i$ is the volume in cell $i$. 

The bulk motion of the gas can also affect whether or not a fluid element is shifted into
the line of an outgoing LW photon.  We account for this in an approximate way with an 
additional component to the Doppler broadening parameter, 
\begin{equation}
b_\mathrm{D,T,v} = \sqrt{b_\mathrm{D,T}^2 + b_\mathrm{D,v}^2} = 
\sqrt{\frac{2 k_B T}{m_{\mathrm{H}_2}} + 2 \sigma^2_\mathrm{mt} },
\end{equation}
where $\sigma_\mathrm{mt}$ is taken to be the microturbulent velocity dispersion, 
which is approximated as the H$_2$-weighted velocity dispersion of the halo 
in analogy to $T_\mathrm{eff}$. 
We then 
calculate $f_\mathrm{shield}^{\mathrm{H}_2}$ with the fitting function in WH11
\begin{align}
f_\mathrm{shield}^{\mathrm{H}_2} (& N_{H_2}, b_\mathrm{D,T,v})  = \frac{0.9379}{(1+x/D_\mathrm{H_2})^{1.879}}   \nonumber \\ 
  + & \frac{0.03465}{(1+x)^{0.473}} \times \exp[-2.293 \times10^{-4}\sqrt{1+x}], 
\end{align}
where $x=N_{\mathrm{H}_2}/(8.465 \times 10^{13} \mathrm{cm}^{-2})$ and $D_{\mathrm{H}_2} 
= b_\mathrm{D,T,v}/(10^5 \mathrm{cm}\ \mathrm{s}^{-1})$.

\begin{figure*}
\includegraphics{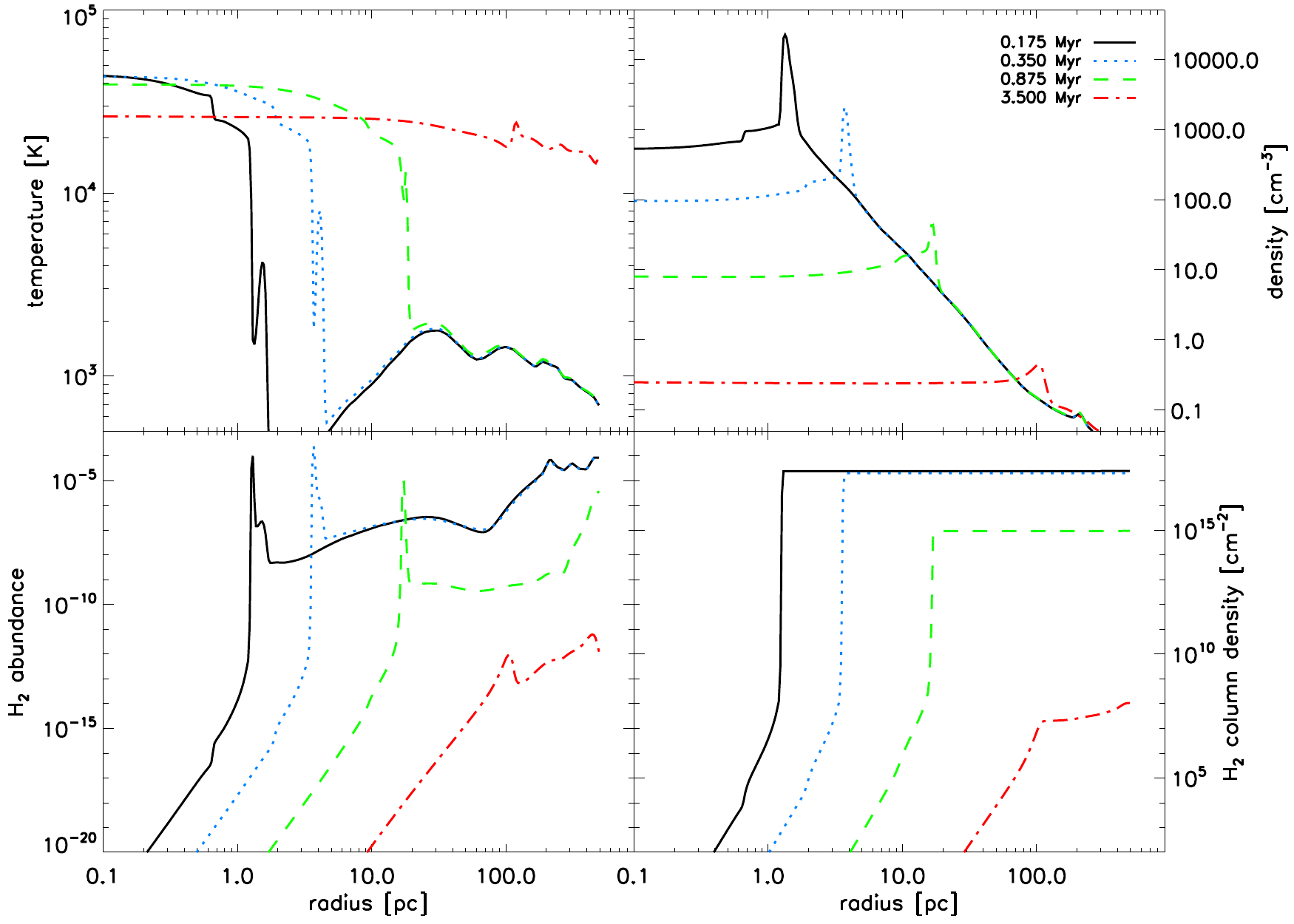}
\caption{Evolution of the I-front of the 60 \Ms\ star in halo 1.  The times are 0.175 Myr
(black solid line), 0.35 Myr (blue dotted line), 0.875 Myr (green dashed line) and 3.5
Myr, the end of the life of the star (red dash-dotted line).}
\label{fig:fesc-r}
\end{figure*}

We construct $f_\mathrm{shield}^\mathrm{H}$ from the fitting function in WH11,
\begin{equation}
f_\mathrm{shield}^\mathrm{H} (N_\mathrm{H}) = \frac{1}{(1+x_\mathrm{H})^{1.62}} 
\exp(-0.149 \, x_\mathrm{H}),
\end{equation}
where $x_\mathrm{H} = N_\mathrm{H}/(2.85 \times 10^{23} \mathrm{cm}^{-2})$.  Escape fractions in the 
near-field limit are calculated for each profile of the H{\sc{ii}} region and surrounding halo as 
they evolve throughout the run, typically 1000 times during the simulation.

\subsubsection{Far-Field Limit}

In the far-field approximation, we begin by calculating the dimensionless equivalent width 
of all the LW lines, $\widetilde W^{\prime}$.  It cannot exceed the total 
dimensionless width of the LW range 11.2 -- 13.6 eV:  $W^\prime_\mathrm{
max} = \ln(1110/912)\simeq 0.2$.  The escape fraction in the far-field limit is then one 
minus the fraction of these two quantities, $f_\mathrm{esc} = 1 - \widetilde W^{\prime}/W^\prime_
\mathrm{max}$. 

The transition from a lower level $l$ (with vibrational quantum number $v$ and 
rotational quantum number $J$) to an upper level $u$ (with $v^\prime$ and $J^
\prime$, respectively) depends on the photoabsorption cross-section $\sigma_{ul}$ and 
column density $N_l$.  In a first step, we show the calculation for a single line. Later, we 
combine them to obtain the total equivalent width.  We follow the notation of DB96.  
The equivalent width of the $l$ to $u$ transition can be written as
\begin{equation}
W_{ul} (N_l)= \int [1-\exp(-N_l \sigma_{ul})] \mathrm{d}\nu .
\end{equation}
If $N_l \sigma_{ul} \ll 1$ we can treat the line width in the weak-line limit, in which 
$\exp(-N_l \sigma_{ul}) \simeq 1 - N_l \sigma_{ul}$. The photodissociation cross section 
can also be written in terms of the cross section at line center, $\sigma_{ul}(\nu_0)$, 
and a frequency dependent line profile function that is normalized to one. Therefore, we 
have
\begin{equation}
W_{ul,\mathrm{W}} (N_l) \simeq N_l \sigma_{ul}(\nu_0).
\end{equation}
The shape of the equivalent width needs to be accounted for in the general case.  Two 
processes determine the shape, Lorentz and Doppler broadening.  

Lorentz broadening, also known as natural broadening, is intrinsic to the line and is a 
consequence of the Uncertainty Principle:  a quantum state with a lifetime $\Delta t$ has 
an uncertainty associated with its energy $\Delta E$ such that $\Delta E \Delta t \geq 
\hbar / 2$. Consequently, radiative transitions to or from this state do not have a precise 
energy but instead occur with a range of energies with a distribution with a width $\sim 
\Delta E$. Doppler broadening, also known as thermal broadening, is a consequence of 
the thermal motion of molecules and atoms.  The Doppler shifts associated with this motion again 
lead to a spread in the frequency of the transition in the laboratory frame.  If the 
gas particles have a Maxwell-Boltzmann velocity distribution, the line profile is a simple 
Gaussian when Doppler broadening dominates. 

Doppler broadening usually dominates near the center of the line $\nu_0$. However, the Doppler 
line profile function falls off exponentially away from the line center while the Lorentz profile 
falls off only as $(\nu - \nu_{0})^{-2}$.  Lorentz broadening therefore always dominates 
far enough from the centre of the line. Lorentz broadening can often be ignored far from 
the centre of the line because the line is so weak there that the optical depth due to that 
region of the line profile is $\ll 1$.  However, if $N_{l}$ is very large this may no longer 
be true and we need to account for both Lorentz and Doppler broadening.

When both effects are important, the line profile becomes a convolution of the Lorentz 
and Doppler profiles known as the Voigt profile.  Instead of calculating the Voigt profile 
directly, we follow the approach of \citet[][hereafter RW74]{rw74} and approximate the 
equivalent width of a single line to be
\begin{equation}
 W_{ul} = [ W_{ul,\mathrm{L}}^2 +  W_{ul,\mathrm{D}}^2 - 
          ( W_{ul,\mathrm{L}}*W_{ul,\mathrm{D}}/W_{ul,\mathrm{W}})^2]^{1/2}
\end{equation}
(equation 3 of RW74).  Here, $W_{\rm L}$ is the equivalent width of the line, assuming 
only Lorentz broadening, $W_{\rm D}$ is the equivalent width for only the Doppler 
effect taken into account, while $W_{\rm W}$ is the equivalent width in the weak line limit (equation 
10). 

We use the approximation in RW74 to calculate $W_{\rm D}$,
\begin{equation}
W_{\rm D} = \Delta \nu_{\rm D} D(z),
\end{equation}
where  $\Delta \nu_{\rm D} = b_\mathrm{D,T,v} \nu_{0} / c$ includes 
the Doppler broadening parameter $b_\mathrm{D,T,v}$, as defined in Equation (6). 
The function $D$ is a seventh-order polynomial given in the Appendix of RW74 
and $z = \sigma_{ul}(\nu_{0}) N_{l} / (\Delta \nu_{\rm D} \pi^{1/2})$.

We take the expression for $W_{\rm L}$ given in \citet{bel00}, which is more accurate than the one from RW74,
\begin{equation}
W_{\rm L} = 2 \pi \Delta \nu_{\rm L} L(z),
\end{equation}
where $\Delta \nu_{\rm L} = \Gamma/2$ is the Lorentz half-width of the line with the total 
de-excitation rate $\Gamma$, $z = \sigma_{ul}(\nu_{0}) N_{l} / (2 \pi \Delta \nu_{\rm L})$, 
and $L$ is the Ladenburg-Reiche function used by \citet{bel00}. To combine the individual 
lines we first find their dimensionless equivalent widths.  As the width of the line is small compared to its 
frequency, it can be written as
\begin{equation}
W_{ul}^\prime =  \frac{W_{ul}}{\nu_0}.
\end{equation}
We assume an ortho-to-para ratio of 3:1 for molecular hydrogen, 
so $N_{\mathrm{ortho}} = 0.75 \times N_{\mathrm{H}_2}$ 
and $N_{\mathrm{para}} = 0.25 \times N_{\mathrm{H}_2}$.  Applying the transition rules, only one 
upper rotational level can be reached: $J^\prime=1$.  We further limit our calculations to 
transition energies below the ionization limit of hydrogen, $E \ge \unit[13.6]{eV}$, because 
we assume that any photon with $E \ge \unit[13.6]{eV}$ ionizes a hydrogen atom. 

For each transition, we take the molecular data required to compute the dimensionless
equivalent width -- the oscillator strength and frequency of the transition, and the total radiative
de-excitation rate of the excited state -- from the papers by \citet{ar89} and \citet{abg92}.

In the case of no line overlap, the total dimensionless equivalent width can be calculated 
by a sum over all individual lines,
\begin{equation}
W^\prime = \sum_l \sum_u W_{ul}^\prime .
\end{equation} 
If there is overlap we account for it in the same way as DB96.  To ensure that the 
total dimensionless equivalent width of our set of lines satisfies this constraint, we write 
it as
\begin{equation}
\tilde{W}^{\prime} = W^{\prime}_{\rm max} \left[1 - \exp(-W^{\prime} / W^{\prime}_{\rm 
max}) \right].
\end{equation}
Therefore, since $f_{\rm esc} = 1 - f_{\rm abs} = 1 - \tilde{W}^{\prime}/{W^{\prime
}_{\rm max}}$, we have
\begin{equation}
f_\mathrm{esc} =  \exp\left( -W^\prime/W^\prime_\mathrm{max}\right).
\end{equation}

As in the near-field case, we consider shielding by neutral hydrogen in addition to self
shielding by H$_2$.  All Lyman lines starting from Ly$\beta$ fall into the LW range and 
can therefore reduce the LW escape fraction.  Like WH11, we consider Lyman 
transitions ($l=1$) from $u=2$ up to $u=10$.   We calculate $W^\prime_\mathrm{H}$ 
with data from \citet{wf09} in analogy to $W^\prime$. The two dimensionless equivalent 
widths are then summed and the total escape fraction becomes
\begin{equation}
f_\mathrm{esc,H_2 \& H} =  \exp\left( -\frac{W^\prime + W^\prime_\mathrm{H}}{W^\prime_\mathrm{max}}\right).
\end{equation}

\section{Results}

We first examine the evolution of the I-front and H$\,${\sc{ii}} region in a halo and then 
present our results for $f_\mathrm{esc}$ in the near-field and far-field limits.

\subsection{Evolution of the H{\sc{ii}} Region}

The evolution of the I-front and H$\,${\sc{ii}} region depends on the mass of the star and its host 
halo.  At early times the I-front propagates very rapidly, leaving the gas behind it essentially
undisturbed (an R-type front). It then decelerates as it approaches the 
Str\"{o}mgren radius of the halo.  If the halo is massive and the star is not very luminous 
the I-front stalls at the Str\"{o}mgren radius and advances no further. If the star is bright, 
the front may briefly loiter at the Str\"{o}mgren radius, but it then resumes its expansion 
driving a shock in front of it (a D-type front). As it descends the steep 
density gradient of the halo, the front can break through the shell and revert to R-type, 
flash ionizing the halo out to radii of 2.5--5 kpc. The I-front may be preceded by an H$_
2$ photodissociation front driven by LW photons (see e.g. studies by \citealt{rgs01,rgs02}).  
This second radiation front is often at 
first confined to the halo because the rate of H$_2$ formation in the photodissociation 
region (PDR) exceeds the rate at which LW photons are emitted by the star.

The hard UV spectra of hot high-mass Pop III stars increases the thickness of the I-front because 
of the larger range of mean free paths of the ionizing photons in the neutral gas.  The 
outer layers of the front can have temperatures of just a few thousand K and free 
electron fractions of $\sim$ 10\%, ideal conditions for the formation of H$_2$ in the gas 
phase via the H$^-$ and H$_2^+$ channels:
\begin{equation}
\mathrm{H} + \mathrm{e}^- \rightarrow \mathrm{H}^- + \gamma \hspace{0.25in} 
\mathrm{H}^- + \mathrm{H} \rightarrow \mathrm{H}_2 + \mathrm{e}^-  \hspace{0.07in} 
\end{equation}
\begin{equation}
\mathrm{H} + \mathrm{H}^+ \rightarrow \mathrm{H}_2^+  + \gamma \hspace{0.25in} 
\mathrm{H}_2^+ + \mathrm{H} \rightarrow \mathrm{H}_2 + \mathrm{H}^+ \vspace{0.1in}\;\;.
\end{equation}
An H$_2$ layer may thus form in the outer shell of the front, with a molecular 
mass that greatly exceeds the one in the surrounding PDR.  In the H{\sc{ii}} region itself nearly all 
H$_2$ is collisionally dissociated by free electrons.  Beyond the I-front much of the 
H$_2$ in the halo that existed before the star was born may have been destroyed by the PDR.  
The total H$_2$ column density in the halo may therefore be dominated by the H$_
2$ sandwiched between the I-front and the shell of gas plowed up by the front after 
becoming D-type.  Since this thin layer may govern LW escape from the halo, it is essential
that it is well resolved in our numerical simulations.

We show in Figure \ref{fig:fesc-r} profiles for the I-front, 
H$\,${\sc{ii}} region and PDR for the 60 \Ms\ star in halo 1 at 
0.175, 0.35, 0.875 and 3.5 Myr, the lifetime of the star.  At 
0.175 Myr it is evident from the temperature plot that the I-front has become D-type:  
an ionized region with a temperature of 4.5 $\times$ 10$^4$ K extends out to the position of the 
front at 1.2 pc, after which the temperature drops to 1400 K, rises to 4000 K at 1.5 pc and 
then falls to below 500 K at 1.7 
pc.  The 4000 K gas is the plowed up, shocked material.  The temperature falls to 
1400 K in the region between the fully ionized gas and the dense shell because of
H$_2$ cooling in the outer layers of the I-front.  The shock has fully detached from 
the front and is at $\sim$ 1.7 pc.  The density spike at 1.3 pc marks the center of 
the plowed up shell, which has ten times the density of the ambient halo.  H$_2$ 
mass fractions reach 10$^{-4}$ between the I-front and the dense shell and then fall 
to 10$^{-7}$ just beyond it, marking the beginning of the PDR. The latter extends out 
to $\sim$ 100 pc, beyond which the H$_2$ mass fractions gradually rise to $\sim$ 
10$^{-4}$, those that were in the halo prior to the birth of the star.  Note the 
extremely low H$_2$ mass fractions within the H$\,${\sc{ii}} region that are due to collisional 
dissociation.

At 0.35 Myr, the I-front has advanced to $\sim$ 4 pc, but remains D-type and continues to 
move behind the shock front. The structures of the PDR beyond the I-front are virtually 
identical to the ones at 0.175 Myr, as shown in the H$_2$ mass fractions. 
Note that at each stage the rise of the H$_2$ column density to its peak value 
coincides with the position of the H$_2$ layer just ahead of the front. The thin layer 
in which most of the H$_{2}$ forms has a peak H$_{2}$ abundance of over
$10^{-4}$, slightly higher than at 0.175~Myr. It is also somewhat wider, but the
peak density in the layer is roughly an order of magnitude lower than at 0.175~Myr.
Together, these factors lead to the H$_2$ column density at 0.35 Myr 
being little changed from its value at 0.175 Myr. 

The H$\,${\sc{ii}} region remains D-type until about 0.875 Myr, when the dense shell is 
barely visible in the temperature profile at $\sim$ 20 pc. The peak H$_{2}$
abundance in the shell at this point has fallen to 10$^{-5}$ because of 
the smaller densities there, resulting in a substantially lower H$_2$ column density. 
As the halo becomes more and more ionized, more LW photons 
pass through the I-front and decrease the H$_2$ abundance in the PDR. 
This also increases the radius of the PDR, from 100 pc at 0.375 Myr to 250 pc at 0.875 Myr.  
The peak density behind the shock front 
has fallen to $\sim$ 75 cm$^{-3}$, but this is still 10 
times higher than the density immediately ahead of the shock.

Later on, the I-front becomes R-type, overruns the PDR and breaks out of the halo.
As the density of the H$_2$ layer falls with the expansion of the I-front, 
the H$_2$ column density decreases. 

At the end of the lifetime of the star, 3.5 Myr, the whole halo is ionized, while the 
shock front is still moving outwards. Nearly all molecular hydrogen is destroyed in the 
H$\,${\sc{ii}} region due to collisional dissociations, resulting 
is a very low H$_2$ column density which is not able to shield against LW radiation. 

This case is an example of delayed breakout, in which the star eventually ionizes 
the halo, but only after a large fraction of its lifetime.  In our grid of models some 
I-fronts loiter at the Str\"{o}mgren radius until the death of the star.  For the 9 \Ms\ star 
in the $1.7 \times 10^7$ \Ms\ halo the Str\"{o}mgren radius is below the grid 
resolution and the H$\,${\sc{ii}} region remains hypercompact until the death of the star.  
In these models the halo is essentially undisturbed by UV radiation from the star.

\subsection{LW Escape Fractions in the Near-Field Limit}

\begin{figure}
\includegraphics[width=0.99\columnwidth]{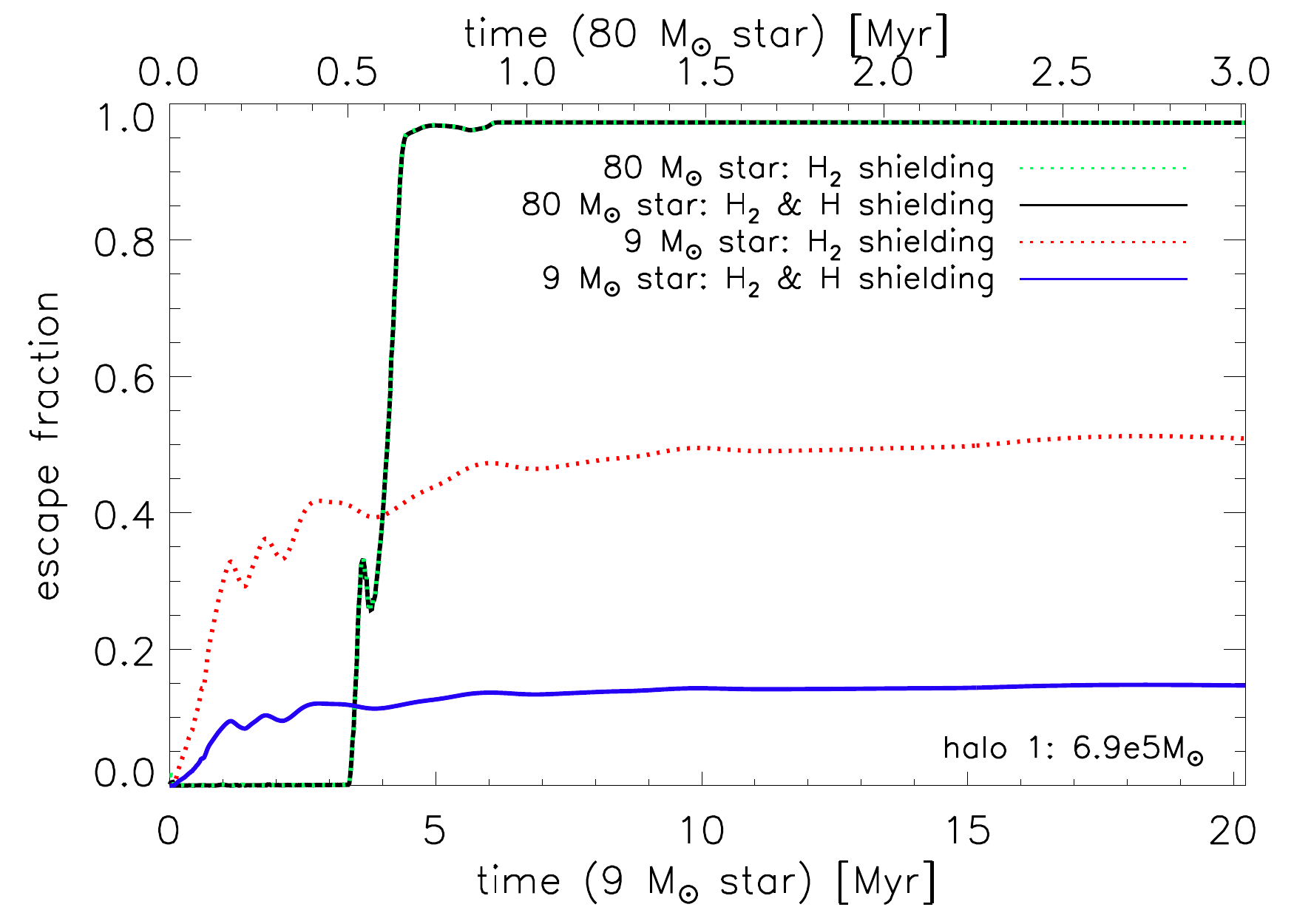}
\caption[Escape fraction as a function of time]
{Smoothed escape fractions for a 9 \Ms\ star (black solid and green dotted lines, overlying) 
and a 80 \Ms\ 
star (blue solid and red dotted lines) in halo 1 as a function of time.  Escape fractions 
with shielding by H$_2$ only (dotted lines) and by H$_2$ and neutral hydrogen (solid 
lines) are shown.} 
\label{fig:fesct}
\end{figure}

The escape fraction evolves as the H$\,${\sc{ii}} region of the star expands in the halo and the H$_2$ 
column density and neutral H fraction change. In Figure \ref{fig:fesct}, we show how $f_\mathrm{esc}$ 
evolves over time with H$_2$ self-shielding alone (dotted lines) and  with both H$_2$ and H shielding 
(solid lines) for the 9 \Ms\ and 80 \Ms\ stars in halo 1.  With the 9 \Ms\ star, the 
I-front is trapped and ionizing UV photons never break out of the halo.  With the 80 
\Ms\ star, the I-front breaks out of the halo and fully ionizes it.  Here, the escape 
fraction is essentially zero until the I-front changes to R-type at 0.6 Myr.  Before this 
transition, the H$_2$ column density  is 
high enough to prevent most LW photons from escaping the halo, due to the dense 
shell driven by the I-front.  After the front
breaks through the dense shell and accelerates down the steep density gradient of
the halo, the column density of H$_{2}$ rapidly decreases, owing to the rapid
decrease in the post-shock density. 
Because the density decreases so rapidly
with radius, most of the H$_2$ is destroyed by the front well before it 
reaches the virial radius of the halo at a time $\sim$ 0.9 Myr.

The rapid initial increase in $f_\mathrm{esc}$ with the 9 \Ms\ star is due to the lower 
luminosity of the star, which creates less H$_2$ in the dense shell plowed up by the 
I-front. The dense shell therefore has a lower H$_{2}$ column density and
hence traps LW photons less effectively than in the case of
the 80 \Ms\ star.  However, the smaller UV 
flux also prevents the I-front from overrunning the entire halo, leading to it 
remaining trapped for the lifetime of the star.  Consequently, the H$_2$
column density in the post-shock shell varies only slightly over the lifetime of the
star, and the LW escape fraction tends towards a roughly constant value. If we
account only for H$_2$ self-shielding, then the LW escape fraction at late times
is around 51\%. On the other hand, if we also account for absorption in the 
Lyman series lines of atomic hydrogen, we recover instead an escape fraction of
around 15\% at late times. We see therefore that
shielding due to neutral H is especially important in this case. This is 
because most of the halo is not ionized, and the presence of atomic hydrogen reduces 
$f_\mathrm{esc}$ by more than two thirds compared to the value based on H$_2$ alone. 
We conclude that shielding by neutral hydrogen cannot be neglected for
stars with $M \sim 10 \,$\Ms\ in cosmological halos.

We list near-field escape fractions that are averaged over the lifetime of the star in Tables 
\ref{tab:fescH2} and \ref{tab:fescH2H}. Table~\ref{tab:fescH2} shows the values 
that we obtain if we only account for H$_{2}$ self-shielding, while Table~\ref{tab:fescH2H}
shows the corresponding values for the case where we account for shielding by both H
and H$_{2}$. We see that on the whole, we recover similar values in both cases, unless
the mass of the star is low, or the mass of the halo is large.

\begin{figure}
\includegraphics[width=0.99\columnwidth]{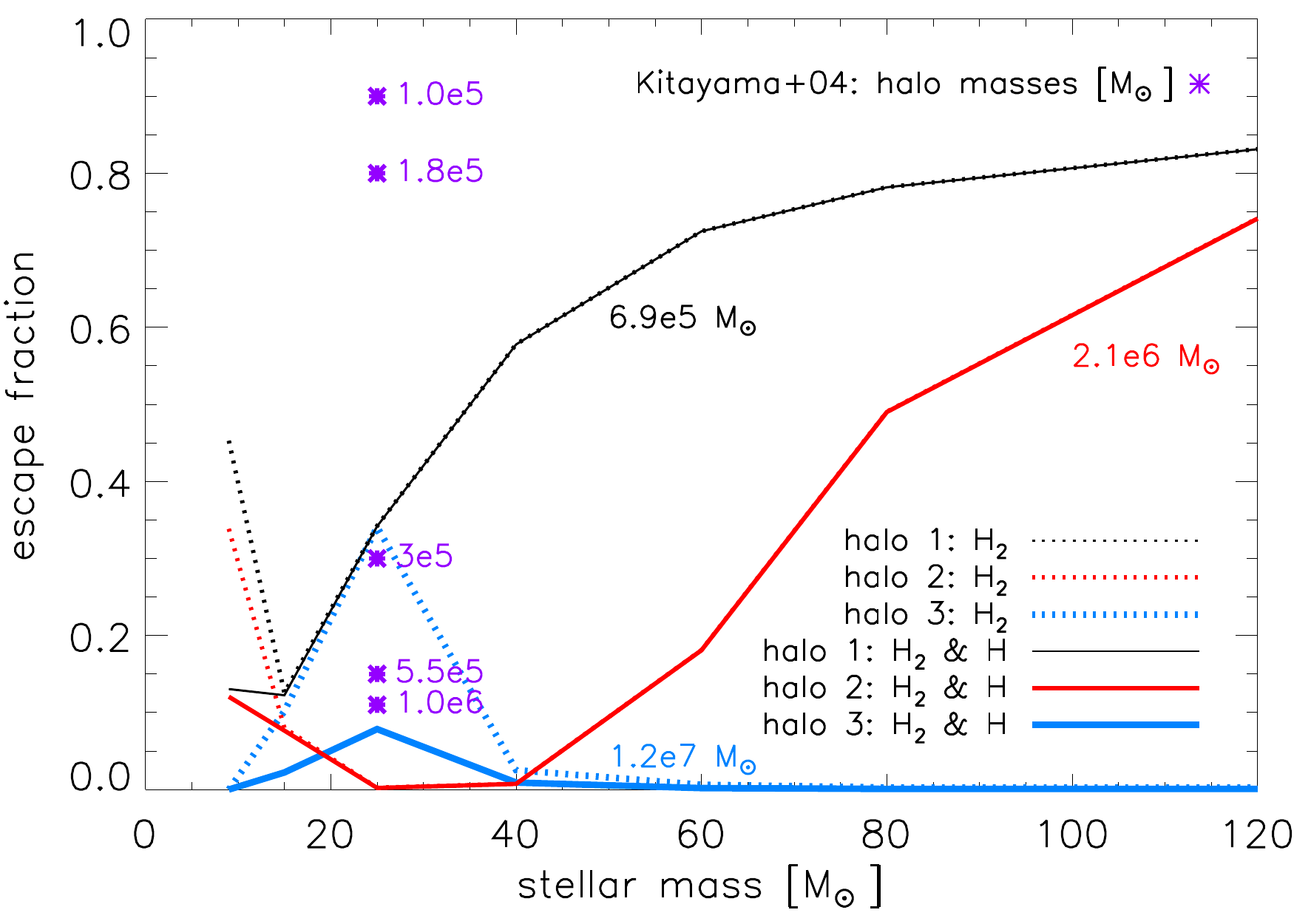}
\caption[Time-averaged escape fractions]{Time-averaged escape fractions for all our
models as a function of stellar mass for halo 1 (black thin line), halo 2 
(red normal line), and halo 3 (blue thick line). 
Escape fractions with just shielding by H$_2$ (dotted lines) and by H$_2$ and neutral 
hydrogen (solid lines) are shown, together with some values from \citet{ket04} (purple 
stars).}
\label{fig:fescave}
\end{figure}

Our time-averaged values for the near-field escape fraction are also plotted for all three 
halos as a function of stellar mass in Figure \ref{fig:fescave}.  For a given star, 
$f_\mathrm{esc}$ falls with increasing halo 
mass because a smaller fraction of the halo becomes ionized.  In halos 1 and 2, $f_\mathrm{esc}$ increases
with stellar mass, because the higher luminosities of the more massive stars 
ionize more of the halo.  None of 
the stars in our study can ionize halo 3, which is not only the most massive halo but 
is also the densest (see Figure \ref{fig:haloics}).  All LW 
photons are also trapped, even for the 120 \Ms\ star.  With more massive stars in 
halo 3, the H$\,${\sc{ii}} region does expand, but the star dies before the I-front can reach the virial 
radius. 

Note that escape fractions for small stars (the 9 \Ms\ star in halo, the 9 \Ms\ 
and 15 \Ms\ stars in halo 2 and the 25 
\Ms\ star in halo 3) can actually be larger than for more massive stars in the same 
halo.  The lower luminosities of these less massive stars produce less H$_2$ in the 
dense shell surrounding the H{\sc{ii}} region so more LW photons can escape the halo.  
We can compare our numbers with the study by \citet{ket04}. However, we note that they 
are using an earlier self-shielding function from DB96 and the overlap in parameter 
space is quite limited. They find the escape fraction for a 25 \Ms\ star in a $5.5 \times 
10^5\,$\Ms~halo to be about half the value for our $6.9 \times 10^5\,$\Ms~halo.  As \citet{ket04} 
consider their escape fractions to be lower limits, and given the differences in halo 
structure between our models and theirs, we conclude that the two approaches yield 
consistent results.

\begin{table}
\centering
\begin{tabular}{c|ccccccc}
M [\Ms] & 9  & 15 & 25 & 40 & 60 & 80 & 120 \\
\hline
halo 1 & 45    & 13 & 34   & 58   & 72    & 78   & 83 \\
halo 2 & 34    & 8  & $<1$ & $<1$ & 18    & 49   & 74 \\
halo 3 &  $<1$ & 10 & 34   & 3    &  $<1$ & $<1$ & $<1$ \\
\end{tabular}
\caption[Escape fractions, H$_2$ self-shielding]{Near-field escape fractions (in percent)
averaged over the lifetime of the stars, accounting only for H$_2$ self-shielding}
\label{tab:fescH2}
\end{table}

\begin{table}
\centering
\begin{tabular}{c|ccccccc}
M [\Ms] & 9  & 15 & 25 & 40 & 60 & 80 & 120 \\
\hline
halo 1 & 13   & 12   & 34   & 58   & 72   & 78   & 83   \\
halo 2 & 12   &  8   & $<1$ & $<1$ & 18   & 49   & 74   \\
halo 3 & $<1$ &  2   &  8   & $<1$ & $<1$ & $<1$ & $<1$ \\
\end{tabular}
\caption[Near-field. Escape fractions, H$_2$ and H shielding]{Near-field escape fractions 
(in percent) averaged over the lifetime of the stars, accounting for shielding by both H$_2$ and H}
\label{tab:fescH2H} 
\end{table}

\subsection{Far-Field Limit}

\begin{figure}
\includegraphics[width=0.99\columnwidth]{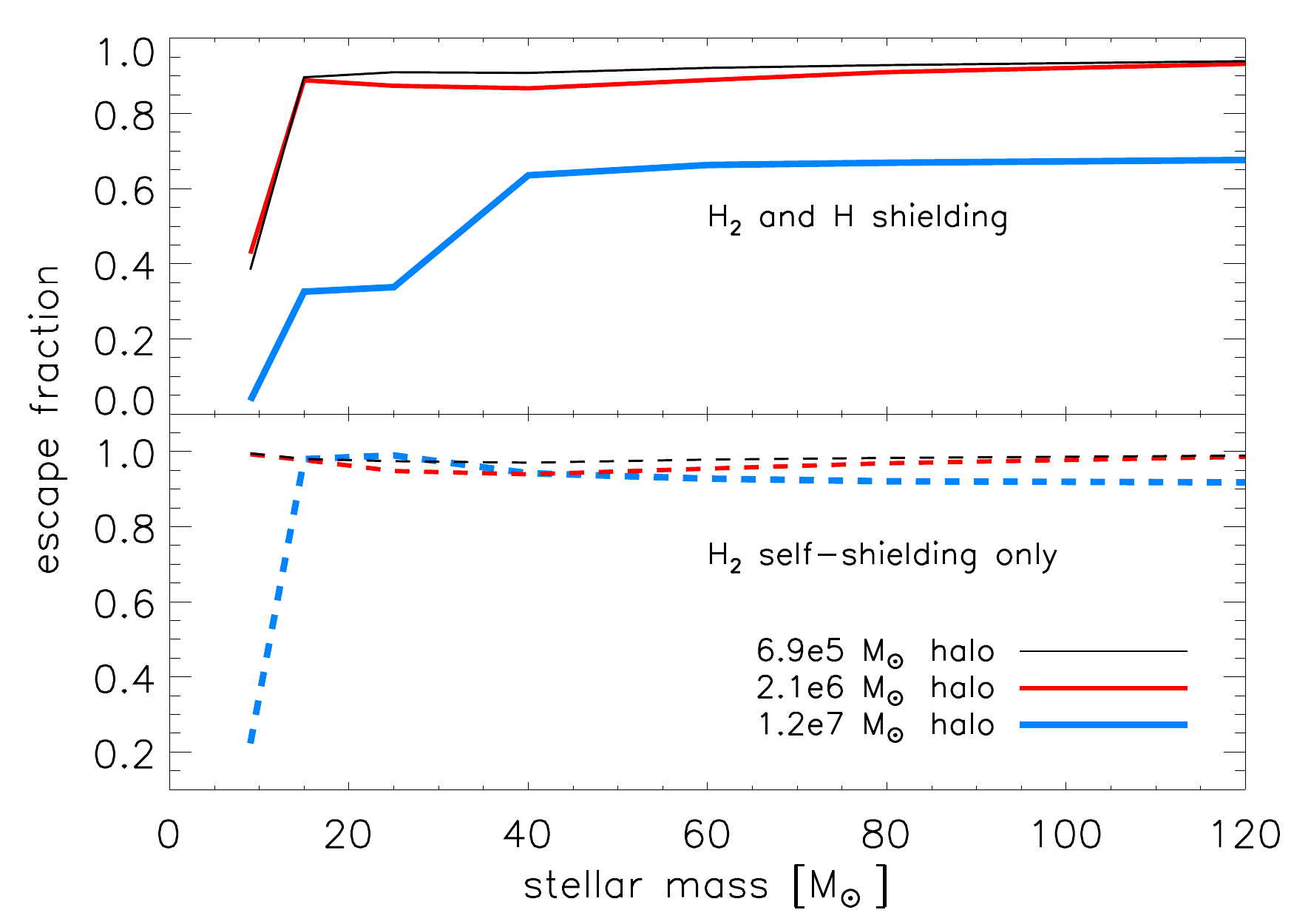}
\caption[Time-averaged escape fractions]{Time-averaged escape fractions in the far
field limit for all the models in our study, plotted as a function of stellar mass (halo 1:  
black thin line, halo 2: red normal line, halo 3: blue thick line). 
In the lower panel, escape fractions for 
self-shielding by H$_2$  alone are shown (dashed lines). In the upper panel, escape 
fractions that account for shielding by H and H$_2$ (solid lines).}
\label{fig:fescfar}
\end{figure}

As expected, escape fractions in the far-field limit are generally much higher than in 
the near-field.  We show time-averaged escape fractions with self-shielding due to H$
_2$ only for all our runs in Table \ref{tab:fescff} and the upper panel of Figure 
\ref{fig:fescfar}, and for shielding by H$_2$ and H in Table \ref{tab:fescffhi} and 
the lower panel of Figure \ref{fig:fescfar}, respectively. 

When considering only shielding by H$_2$, 
for halos 1 and 2, the escape fraction is always higher than 90\% and 
nearly all LW photons escape the halo.
In halo 3, most of the LW radiation from the 9 
\Ms\ star is trapped in the halo, and only a quarter of the photons can escape. The higher 
mass stars all have larger luminosities so their radiation eventually escapes halo 3.

\begin{table}
\centering
\begin{tabular}{c|ccccccc}
M [\Ms] & 9  & 15 & 25 & 40 & 60 & 80 & 120 \\
\hline
halo 1 & 99   & 98 & 97 & 97 & 98 & 98 & 99 \\
halo 2 & 99   & 98 & 95 & 94 & 95 & 97 & 99 \\
halo 3 & 22   & 98 & 99 & 94 & 93 & 92 & 92 \\
\end{tabular}
\caption[Escape fractions, far-field]{Far-field escape fractions (in percent) averaged over 
the lifetime of the star, accounting only for H$_2$ self-shielding.}
\label{tab:fescff}
\end{table}

\begin{table}
\centering
\begin{tabular}{c|ccccccc}
M [\Ms] & 9  & 15 & 25 & 40 & 60 & 80 & 120 \\
\hline
 halo 1 & 38   & 90   & 91   & 91   & 92   & 93   & 94   \\
 halo 2 & 43   & 89   & 87   & 87   & 89   & 91   & 93   \\
 halo 3 &  4   & 33   & 34   & 64   & 66   & 67   & 68   \\
\end{tabular}
\caption[Escape fractions, far-field]{Far-field escape fractions averaged over the lifetime 
of the stars, accounting for shielding by both H$_2$ and H.}
\label{tab:fescffhi}
\end{table}

As in the near-field limit, including shielding by atomic hydrogen reduces the escape fraction.
Less than half of the LW photons from the 9 \Ms\ star escape any halo. 
In particular the escape fraction in halo 3 is only $f_\mathrm{esc} \approx 4$\%.  
LW escape fractions in halos 1 and 2 for more massive 
stars are above 85\%, and they rise slightly with halo mass.  In halo 3, the picture is 
different: for the 9, 15 and 25 \Ms\ stars, the I-front is static and ionizing photons are 
trapped in the halo. The Str\"{o}mgren radii of the I-fronts of the 15 and 25 \Ms\ stars are 
similar, 0.13 pc and 0.14 pc respectively, which explains the step in escape fraction over 
that mass range.  With more massive stars the I-front eventually advances beyond the 
Str\"{o}mgren radius, but the star dies before it can reach the virial radius.  
The final radii of the H$\,${\sc{ii}} regions increase only slightly with the mass of the 
star, because although they are more luminous 
these stars also have shorter lives. The  radii vary from 12 to 20 pc for 40--120 \Ms\ 
stars, resulting in escape fractions of 64\% to 68\%.

\section{Conclusion}

We have calculated LW escape fractions for 9--120 \Ms\ Pop III stars 
in three different primordial halos ranging in mass from $6.9 \times 10^5$ to 
$1.2 \times 10^7$ \Ms. We have considered two limiting cases. The near-field limit 
is relevant for objects close to the halo that do not have a significant velocity
relative to it, so that the LW absorption lines in the rest-frame of the object coincide
with those in the rest-frame of the halo.
In this limit, the escape fraction can be approximated by the product of the self-shielding 
function of H$_2$  and the shielding function of H$_2$ by neutral hydrogen 
which have been developed by WH11. 
The far-field limit is valid for objects that are significantly red-shifted or blue-shifted with 
respect to the halo, usually at cosmological distances. In this limit, we can
estimate the escape fraction by calculating
the ratio of the equivalent width of the LW absorption lines 
to the width of the entire LW frequency range. 

In both limits, we consider self-shielding by molecular hydrogen alone as well as 
the combined effects of molecular and atomic gas together. We find that it is 
important to consider both types of shielding. 

In the near-field, the escape fraction generally rises with increasing mass of the central star. 
In the $6.9 \times 10^5$ \Ms\ halo, it grows from about 10\% to about 85\% and for the $2.1 \times 10^6$ \Ms\ 
halo to about 75\%. None of the stars in our study are bright enough to fully ionize 
the $1.2 \times 10^7$ \Ms\ halo and escape fractions from this halo are nearly zero in the near-field. 
For some stars with masses close to 10 \Ms, escape fractions can be higher than for slightly more 
massive stars because their lower luminosities cannot build up as large an H$_2$ column
density, e.g. 12\% of the LW photons are able to escape the $2.1 \times 10^6$ \Ms\ halo for a 
9 \Ms\ star, but only 8\% for a 15 \Ms\ star. 

In the far-field, the escape fractions are generally higher. 
Only in the case of the 9 \Ms\ star do fewer than 50\% of its LW photons exit all of our halos. In the most massive 
halo, this is also true for the 15 and 25 \Ms\ stars. For higher stellar masses, the escape fraction rises 
to above 90\% for the less massive halos and to above 60\% for the $1.2 \times 10^7\,$\Ms$\,$ halo.  

If shielding by neutral H is neglected, the escape fractions are overpredicted. 
Therefore, the additional shielding function of neutral hydrogen from WH11 provides an important 
improvement to the older model from DB96. 
This is especially 
severe for stars with masses up to 25 \Ms\ in the near-field, where shielding by neutral H reduces 
the escape fraction by as much as a factor of three.
In the far-field, less than 50\% of the LW radiation
from a 9 \Ms\ star can exit any halo, but if neutral hydrogen shielding is neglected up to 99\% can 
exit the halo. The $1.2 \times 10^7\,$\Ms$\,$  halo is not ionized by 
any of the central stars considered, so 
its large neutral hydrogen column density plays a major role in shielding, reducing the escape 
fraction from about 90\% to about 60\% for stars with masses of 40 \Ms\ and higher. 

Our models assume 1D spherical symmetry.  In reality, both ionizing and LW photons can
break out of the halo along lines of sight with lower optical depths, so our escape fractions
should be taken to be lower limits.  We note in particular that shielding by neutral H, which 
has largely been ignored in studies until now, can play a large role in LW escape fractions, 
especially those of lower-mass Pop III stars.  This latter point is important because of the
relative contributions of lower-mass and higher-mass Pop III stars to the LW background.  It
has recently been found that more realistic stellar spectra that contain a lower-mass Pop III
star component may be less efficient at photodissociating H$_2$ in primordial halos than
previously thought \citep{agar15}.  Shielding by H in the host halos of such lower-mass Pop
III stars may reinforce this trend by allowing fewer LW photons to escape into the cosmos.

Finally, we note that star formation in the vicinity of a halo hosting a low-mass Pop III star
may be less influenced by the star than by the global LW background because of low 
escape fractions from the halo in the near field.  Conversely, smaller LW escape fractions 
from low-mass Pop III stars have less of an effect on the LW background on cosmological 
scales because escape fractions in the far-field limit are generally much larger.

\section*{Acknowledgments}

We thank the anonymous referee for a careful review of our paper. 
We thank Volker Bromm, Andreas Burkert, Andrea Ferrara, Tilman Hartwig
and Jan-Pieter Paardekooper for useful discussions.  
The authors acknowledge support 
from the European Research Council under the European Community's Seventh Framework 
Programme (FP7/2007 - 2013) via the ERC Advanced Grant ``STARLIGHT: Formation of the 
First Stars" (project number 339177). SCOG and RSK also acknowledge support from 
the Deutsche Forschungsgemeinschaft via SFB 881, ``The Milky Way System'' (sub-projects
B1, B2 and B8) and SPP 1573 , ``Physics of the Interstellar Medium'' (grant number GL 668/2-1).

\bibliographystyle{mn2e}

\begin{thebibliography}{84}
\expandafter\ifx\csname natexlab\endcsname\relax\def\natexlab#1{#1}\fi

\bibitem[{{Abel}, {Bryan} \& {Norman}(2000){Abel}, {Bryan}, \&
  {Norman}}]{abn00}
{Abel} T., {Bryan} G.~L., {Norman} M.~L., 2000, \apj, 540, 39

\bibitem[{{Abel}, {Bryan} \& {Norman}(2002){Abel}, {Bryan}, \&
  {Norman}}]{abn02}
{Abel} T., {Bryan} G.~L., {Norman} M.~L., 2002, Science, 295, 93

\bibitem[{{Abel}, {Wise} \& {Bryan}(2007){Abel}, {Wise}, \& {Bryan}}]{awb07}
{Abel} T., {Wise} J.~H., {Bryan} G.~L., 2007, \apj, 659, L87

\bibitem[{{Abgrall} {et~al}\mbox{.}(1992){Abgrall}, {Le Bourlot}, {Pineau Des
  Forets}, {Roueff}, {Flower}, \& {Heck}}]{abg92}
{Abgrall} H., {Le Bourlot} J., {Pineau Des Forets} G., {Roueff} E., {Flower}
  D.~R., {Heck} L., 1992, \aap, 253, 525

\bibitem[{{Abgrall} \& {Roueff}(1989)}]{ar89}
{Abgrall} H., {Roueff} E., 1989, \aaps, 79, 313

\bibitem[{{Agarwal} \& {Khochfar}(2015)}]{agar15}
{Agarwal} B., {Khochfar} S., 2015, \mnras, 446, 160

\bibitem[{{Alvarez}, {Bromm} \& {Shapiro}(2006){Alvarez}, {Bromm}, \&
  {Shapiro}}]{abs06}
{Alvarez} M.~A., {Bromm} V., {Shapiro} P.~R., 2006, \apj, 639, 621

\bibitem[{{Anninos} {et~al}\mbox{.}(1997){Anninos}, {Zhang}, {Abel}, \&
  {Norman}}]{anet97}
{Anninos} P., {Zhang} Y., {Abel} T., {Norman} M.~L., 1997, New Astronomy, 2,
  209

\bibitem[{{Beers} \& {Christlieb}(2005)}]{bc05}
{Beers} T.~C., {Christlieb} N., 2005, \araa, 43, 531

\bibitem[{{Belafhal}(2000)}]{bel00}
{Belafhal} A., 2000, \optcomm, 177, 111

\bibitem[{{Bromm}, {Coppi} \& {Larson}(1999){Bromm}, {Coppi}, \&
  {Larson}}]{bcl99}
{Bromm} V., {Coppi} P.~S., {Larson} R.~B., 1999, \apj, 527, L5

\bibitem[{{Bromm}, {Coppi} \& {Larson}(2002){Bromm}, {Coppi}, \&
  {Larson}}]{bcl02}
{Bromm} V., {Coppi} P.~S., {Larson} R.~B., 2002, \apj, 564, 23

\bibitem[{Bryan {et~al}\mbox{.}(2014)Bryan {et~al.}}]{bryan14}
Bryan G.~L., {et~al.}, 2014, \apjs, 211, 19

\bibitem[{{Ciardi} \& {Ferrara}(2005)}]{cf05}
{Ciardi} B., {Ferrara} A., 2005, \ssr, 116, 625

\bibitem[{{Clark} {et~al}\mbox{.}(2011a){Clark}, {Glover}, {Klessen}, \&
  {Bromm}}]{cgkb11}
{Clark} P.~C., {Glover} S.~C.~O., {Klessen} R.~S., {Bromm} V., 2011a, \apj,
  727, 110

\bibitem[{{Clark} {et~al}\mbox{.}(2011b){Clark}, {Glover}, {Smith}, {Greif},
  {Klessen}, \& {Bromm}}]{clark11}
{Clark} P.~C., {Glover} S.~C.~O., {Smith} R.~J., {Greif} T.~H., {Klessen}
  R.~S., {Bromm} V., 2011b, Science, 331, 1040

\bibitem[{{Dopcke} {et~al}\mbox{.}(2013){Dopcke}, {Glover}, {Clark}, \&
  {Klessen}}]{dgck13}
{Dopcke} G., {Glover} S.~C.~O., {Clark} P.~C., {Klessen} R.~S., 2013, \apj,
  766, 103

\bibitem[{{Draine} \& {Bertoldi}(1996)}]{db96}
{Draine} B.~T., {Bertoldi} F., 1996, \apj, 468, 269

\bibitem[{{Frebel}(2010)}]{frebel10}
{Frebel} A., 2010, Astronomische Nachrichten, 331, 474

\bibitem[{{Frebel}, {Johnson} \& {Bromm}(2008){Frebel}, {Johnson}, \&
  {Bromm}}]{fjb08}
{Frebel} A., {Johnson} J.~L., {Bromm} V., 2008, in IAU Symposium, Vol. 255, IAU
  Symposium, {Hunt} L.~K., {Madden} S.~C., {Schneider} R., eds., pp. 336--340

\bibitem[{{Frebel} {et~al}\mbox{.}(2005){Frebel} {et~al.}}]{fet05}
{Frebel} A., {et~al.}, 2005, \nat, 434, 871

\bibitem[{{Glover}(2005)}]{glover05}
{Glover} S., 2005, \ssr, 117, 445

\bibitem[{{Glover}(2013)}]{glov13}
{Glover} S., 2013, in Astrophysics and Space Science Library, Vol. 396,
  Astrophysics and Space Science Library, {Wiklind} T., {Mobasher} B., {Bromm}
  V., eds., p. 103

\bibitem[{{Glover} \& {Brand}(2001)}]{gb01}
{Glover} S.~C.~O., {Brand} P.~W.~J.~L., 2001, \mnras, 321, 385

\bibitem[{{Glover} \& {Brand}(2003)}]{gb03}
{Glover} S.~C.~O., {Brand} P.~W.~J.~L., 2003, \mnras, 340, 210

\bibitem[{{Greif} {et~al}\mbox{.}(2012){Greif}, {Bromm}, {Clark}, {Glover},
  {Smith}, {Klessen}, {Yoshida}, \& {Springel}}]{get12}
{Greif} T.~H., {Bromm} V., {Clark} P.~C., {Glover} S.~C.~O., {Smith} R.~J.,
  {Klessen} R.~S., {Yoshida} N., {Springel} V., 2012, \mnras, 424, 399

\bibitem[{{Greif} {et~al}\mbox{.}(2010){Greif}, {Glover}, {Bromm}, \&
  {Klessen}}]{get10}
{Greif} T.~H., {Glover} S.~C.~O., {Bromm} V., {Klessen} R.~S., 2010, \apj, 716,
  510

\bibitem[{{Greif} {et~al}\mbox{.}(2011){Greif}, {Springel}, {White}, {Glover},
  {Clark}, {Smith}, {Klessen}, \& {Bromm}}]{get11}
{Greif} T.~H., {Springel} V., {White} S.~D.~M., {Glover} S.~C.~O., {Clark}
  P.~C., {Smith} R.~J., {Klessen} R.~S., {Bromm} V., 2011, \apj, 737, 75

\bibitem[{{Haiman}, {Abel} \& {Rees}(2000){Haiman}, {Abel}, \& {Rees}}]{har00}
{Haiman} Z., {Abel} T., {Rees} M.~J., 2000, \apj, 534, 11

\bibitem[{{Haiman}, {Rees} \& {Loeb}(1997){Haiman}, {Rees}, \& {Loeb}}]{hrl97}
{Haiman} Z., {Rees} M.~J., {Loeb} A., 1997, \apj, 476, 458

\bibitem[{{Hartwig} {et~al}\mbox{.}(2015){Hartwig}, {Bromm}, {Klessen}, \&
  {Glover}}]{hb15}
{Hartwig} T., {Bromm} V., {Klessen} R.~S., {Glover} S.~C.~O., 2015, \mnras,
  447, 3892

\bibitem[{{Hasegawa}, {Umemura} \& {Susa}(2009){Hasegawa}, {Umemura}, \&
  {Susa}}]{hus09}
{Hasegawa} K., {Umemura} M., {Susa} H., 2009, \mnras, 395, 1280

\bibitem[{{Hirano} {et~al}\mbox{.}(2015){Hirano}, {Hosokawa}, {Yoshida},
  {Omukai}, \& {Yorke}}]{hir15}
{Hirano} S., {Hosokawa} T., {Yoshida} N., {Omukai} K., {Yorke} H.~W., 2015,
  \mnras, 448, 568

\bibitem[{{Hirano} {et~al}\mbox{.}(2014){Hirano}, {Hosokawa}, {Yoshida},
  {Umeda}, {Omukai}, {Chiaki}, \& {Yorke}}]{hir14}
{Hirano} S., {Hosokawa} T., {Yoshida} N., {Umeda} H., {Omukai} K., {Chiaki} G.,
  {Yorke} H.~W., 2014, \apj, 781, 60

\bibitem[{{Hosokawa} {et~al}\mbox{.}(2011){Hosokawa}, {Omukai}, {Yoshida}, \&
  {Yorke}}]{hos11}
{Hosokawa} T., {Omukai} K., {Yoshida} N., {Yorke} H.~W., 2011, Science, 334,
  1250

\bibitem[{{Hosokawa} {et~al}\mbox{.}(2012){Hosokawa}, {Yoshida}, {Omukai}, \&
  {Yorke}}]{hos12}
{Hosokawa} T., {Yoshida} N., {Omukai} K., {Yorke} H.~W., 2012, \apj, 760, L37

\bibitem[{{Iliev}, {Shapiro} \& {Raga}(2005){Iliev}, {Shapiro}, \&
  {Raga}}]{il05}
{Iliev} I.~T., {Shapiro} P.~R., {Raga} A.~C., 2005, \mnras, 361, 405

\bibitem[{{Jeon} {et~al}\mbox{.}(2012){Jeon}, {Pawlik}, {Greif}, {Glover},
  {Bromm}, {Milosavljevi{\'c}}, \& {Klessen}}]{jeon11}
{Jeon} M., {Pawlik} A.~H., {Greif} T.~H., {Glover} S.~C.~O., {Bromm} V.,
  {Milosavljevi{\'c}} M., {Klessen} R.~S., 2012, \apj, 754, 34

\bibitem[{{Joggerst} {et~al}\mbox{.}(2010){Joggerst}, {Almgren}, {Bell},
  {Heger}, {Whalen}, \& {Woosley}}]{jet10}
{Joggerst} C.~C., {Almgren} A., {Bell} J., {Heger} A., {Whalen} D., {Woosley}
  S.~E., 2010, \apj, 709, 11

\bibitem[{{Johnson} {et~al}\mbox{.}(2009){Johnson}, {Greif}, {Bromm},
  {Klessen}, \& {Ippolito}}]{jlj09}
{Johnson} J.~L., {Greif} T.~H., {Bromm} V., {Klessen} R.~S., {Ippolito} J.,
  2009, \mnras, 399, 37

\bibitem[{{Kitayama} {et~al}\mbox{.}(2004){Kitayama}, {Yoshida}, {Susa}, \&
  {Umemura}}]{ket04}
{Kitayama} T., {Yoshida} N., {Susa} H., {Umemura} M., 2004, \apj, 613, 631

\bibitem[{{Latif} {et~al}\mbox{.}(2014){Latif}, {Schleicher}, {Bovino},
  {Grassi}, \& {Spaans}}]{latif14}
{Latif} M.~A., {Schleicher} D.~R.~G., {Bovino} S., {Grassi} T., {Spaans} M.,
  2014, \apj, 792, 78

\bibitem[{{Machacek}, {Bryan} \& {Abel}(2001){Machacek}, {Bryan}, \&
  {Abel}}]{met01}
{Machacek} M.~E., {Bryan} G.~L., {Abel} T., 2001, \apj, 548, 509

\bibitem[{{Machacek}, {Bryan} \& {Abel}(2003){Machacek}, {Bryan}, \&
  {Abel}}]{mba03}
{Machacek} M.~E., {Bryan} G.~L., {Abel} T., 2003, \mnras, 338, 273

\bibitem[{{McKee} \& {Tan}(2008)}]{tm08}
{McKee} C.~F., {Tan} J.~C., 2008, \apj, 681, 771

\bibitem[{{Nakamura} \& {Umemura}(2001)}]{nu01}
{Nakamura} F., {Umemura} M., 2001, \apj, 548, 19

\bibitem[{{Navarro}, {Frenk} \& {White}(1997){Navarro}, {Frenk}, \&
  {White}}]{nfw97}
{Navarro} J.~F., {Frenk} C.~S., {White} S.~D.~M., 1997, \apj, 490, 493

\bibitem[{{O'Shea} \& {Norman}(2007)}]{on07}
{O'Shea} B.~W., {Norman} M.~L., 2007, \apj, 654, 66

\bibitem[{{Pawlik}, {Milosavljevi{\'c}} \& {Bromm}(2013){Pawlik},
  {Milosavljevi{\'c}}, \& {Bromm}}]{pmb12}
{Pawlik} A.~H., {Milosavljevi{\'c}} M., {Bromm} V., 2013, \apj, 767, 59

\bibitem[{{Ricotti}, {Gnedin} \& {Shull}(2001){Ricotti}, {Gnedin}, \&
  {Shull}}]{rgs01}
{Ricotti} M., {Gnedin} N.~Y., {Shull} J.~M., 2001, \apj, 560, 580

\bibitem[{{Ricotti}, {Gnedin} \& {Shull}(2002){Ricotti}, {Gnedin}, \&
  {Shull}}]{rgs02}
{Ricotti} M., {Gnedin} N.~Y., {Shull} J.~M., 2002, \apj, 575, 49

\bibitem[{{Rodgers} \& {Williams}(1974)}]{rw74}
{Rodgers} C.~D., {Williams} A.~P., 1974, \jqsrt, 14, 319

\bibitem[{{Rydberg}, {Zackrisson} \& {Scott}(2010){Rydberg}, {Zackrisson}, \&
  {Scott}}]{rz10}
{Rydberg} C.~E., {Zackrisson} E., {Scott} P., 2010, in Cosmic Radiation Fields:
  Sources in the early Universe (CRF 2010), {M.~Raue, T.~Kneiske, D.~Horns,
  D.~Elsaesser, \& P.~Hauschildt }, ed., p.~26

\bibitem[{{Safranek-Shrader} {et~al}\mbox{.}(2012){Safranek-Shrader},
  {Agarwal}, {Federrath}, {Dubey}, {Milosavljevi{\'c}}, \& {Bromm}}]{saf12}
{Safranek-Shrader} C., {Agarwal} M., {Federrath} C., {Dubey} A.,
  {Milosavljevi{\'c}} M., {Bromm} V., 2012, \mnras, 426, 1159

\bibitem[{{Salvadori}, {Schneider} \& {Ferrara}(2007){Salvadori}, {Schneider},
  \& {Ferrara}}]{ssa07}
{Salvadori} S., {Schneider} R., {Ferrara} A., 2007, \mnras, 381, 647

\bibitem[{{Schaerer}(2002)}]{schae02}
{Schaerer} D., 2002, \aap, 382, 28

\bibitem[{{Shapiro}, {Iliev} \& {Raga}(2004){Shapiro}, {Iliev}, \&
  {Raga}}]{il04}
{Shapiro} P.~R., {Iliev} I.~T., {Raga} A.~C., 2004, \mnras, 348, 753

\bibitem[{{Smidt} {et~al}\mbox{.}(2014){Smidt}, {Whalen}, {Chatzopoulos},
  {Wiggins}, {Chen}, {Kozyreva}, \& {Even}}]{smidt14a}
{Smidt} J., {Whalen} D.~J., {Chatzopoulos} E., {Wiggins} B.~K., {Chen} K.-J.,
  {Kozyreva} A., {Even} W., 2014, ArXiv e-prints

\bibitem[{{Smith} {et~al}\mbox{.}(2011){Smith}, {Glover}, {Clark}, {Greif}, \&
  {Klessen}}]{sm11}
{Smith} R.~J., {Glover} S.~C.~O., {Clark} P.~C., {Greif} T., {Klessen} R.~S.,
  2011, \mnras, 414, 3633

\bibitem[{{Smith} {et~al}\mbox{.}(2012){Smith}, {Hosokawa}, {Omukai}, {Glover},
  \& {Klessen}}]{shogk12}
{Smith} R.~J., {Hosokawa} T., {Omukai} K., {Glover} S.~C.~O., {Klessen} R.~S.,
  2012, \mnras, 424, 457

\bibitem[{{Stacy} \& {Bromm}(2014)}]{sb14}
{Stacy} A., {Bromm} V., 2014, \apj, 785, 73

\bibitem[{{Stacy}, {Greif} \& {Bromm}(2010){Stacy}, {Greif}, \&
  {Bromm}}]{stacy10}
{Stacy} A., {Greif} T.~H., {Bromm} V., 2010, \mnras, 403, 45

\bibitem[{{Stacy}, {Greif} \& {Bromm}(2012){Stacy}, {Greif}, \&
  {Bromm}}]{stacy12}
{Stacy} A., {Greif} T.~H., {Bromm} V., 2012, \mnras, 422, 290

\bibitem[{{Stecher} \& {Williams}(1967)}]{sw67}
{Stecher} T.~P., {Williams} D.~A., 1967, \apjl, 149, L29

\bibitem[{{Susa}(2007)}]{su07}
{Susa} H., 2007, \apj, 659, 908

\bibitem[{{Susa} \& {Umemura}(2006)}]{su06}
{Susa} H., {Umemura} M., 2006, \apj, 645, L93

\bibitem[{{Susa}, {Umemura} \& {Hasegawa}(2009){Susa}, {Umemura}, \&
  {Hasegawa}}]{suh09}
{Susa} H., {Umemura} M., {Hasegawa} K., 2009, \apj, 702, 480

\bibitem[{{Tumlinson}(2006)}]{tum06}
{Tumlinson} J., 2006, \apj, 641, 1

\bibitem[{{Turk}, {Abel} \& {O'Shea}(2009){Turk}, {Abel}, \& {O'Shea}}]{turk09}
{Turk} M.~J., {Abel} T., {O'Shea} B., 2009, Science, 325, 601

\bibitem[{{Visbal} {et~al}\mbox{.}(2014){Visbal}, {Haiman}, {Terrazas},
  {Bryan}, \& {Barkana}}]{vis14}
{Visbal} E., {Haiman} Z., {Terrazas} B., {Bryan} G.~L., {Barkana} R., 2014,
  \mnras, 445, 107
\bibitem[{{Whalen}, {Abel} \& {Norman}(2004){Whalen}, {Abel}, \&
  {Norman}}]{wan04}
{Whalen} D., {Abel} T., {Norman} M.~L., 2004, \apj, 610, 14

\bibitem[{{Whalen}, {Hueckstaedt} \& {McConkie}(2010){Whalen}, {Hueckstaedt},
  \& {McConkie}}]{wet10}
{Whalen} D., {Hueckstaedt} R.~M., {McConkie} T.~O., 2010, \apj, 712, 101

\bibitem[{{Whalen} \& {Norman}(2006)}]{wn06}
{Whalen} D., {Norman} M.~L., 2006, \apjs, 162, 281

\bibitem[{{Whalen} \& {Norman}(2008{\natexlab{a}})}]{wn08b}
{Whalen} D., {Norman} M.~L., 2008{\natexlab{a}}, \apj, 673, 664

\bibitem[{{Whalen} {et~al}\mbox{.}(2008){Whalen}, {O'Shea}, {Smidt}, \&
  {Norman}}]{wet08b}
{Whalen} D., {O'Shea} B.~W., {Smidt} J., {Norman} M.~L., 2008, \apj, 679, 925

\bibitem[{{Whalen} \& {Norman}(2008{\natexlab{b}})}]{wn08a}
{Whalen} D.~J., {Norman} M.~L., 2008{\natexlab{b}}, \apj, 672, 287

\bibitem[{{Whalen} {et~al}\mbox{.}(2014){Whalen}, {Smidt}, {Even}, {Woosley},
  {Heger}, {Stiavelli}, \& {Fryer}}]{wet13d}
{Whalen} D.~J., {Smidt} J., {Even} W., {Woosley} S.~E., {Heger} A., {Stiavelli}
  M., {Fryer} C.~L., 2014, \apj, 781, 106

\bibitem[{{Whalen} {et~al}\mbox{.}(2013{\natexlab{a}}){Whalen}
  {et~al.}}]{wet12b}
{Whalen} D.~J., {et~al.}, 2013{\natexlab{a}}, \apj, 777, 110

\bibitem[{{Whalen} {et~al}\mbox{.}(2013{\natexlab{b}}){Whalen}
  {et~al.}}]{wet12c}
{Whalen} D.~J., {et~al.}, 2013{\natexlab{b}}, \apj, 768, 95

\bibitem[{{Wise} \& {Abel}(2007)}]{wa07}
{Wise} J.~H., {Abel} T., 2007, \apj, 671, 1559

\bibitem[{{Wise} \& {Abel}(2008)}]{wa08b}
{Wise} J.~H., {Abel} T., 2008, \apj, 685, 40

\bibitem[{{Wise} {et~al}\mbox{.}(2012){Wise}, {Turk}, {Norman}, \&
  {Abel}}]{wise12}
{Wise} J.~H., {Turk} M.~J., {Norman} M.~L., {Abel} T., 2012, \apj, 745, 50

\bibitem[{{Wise} \& {Fuhr}(2009)}]{wf09}
{Wise} W.~L., {Fuhr} J.~R., 2009, \jpcrd, 38, 565

\bibitem[{{Wolcott-Green} \& {Haiman}(2011)}]{wh11}
{Wolcott-Green} J., {Haiman} Z., 2011, \mnras, 412, 2603

\end{thebibliography}
\setlength{\bibhang}{2.0em}
\setlength\labelwidth{0.0em}

\label{lastpage}

\end{document}